\documentclass[11pt,twoside]{article}

\usepackage{asp2006}
\usepackage{epsf}
\usepackage{psfig}
\usepackage{lscape}

\begin{document}
\title{Imaging a Jet Base --- Prospects with M87}

\author{R. Craig Walker}
\affil{National Radio Astronomy Observatory, P. O. Box O, Socorro, NM 
87801 USA}

\author{Chun Ly}
\affil{Department of Astronomy, University of California, Los Angeles,
Los Angeles, CA 90095 USA} 

\author{William Junor}
\affil{ISR-2, MS-D436, Los Alamos National Laboratory, Los Alamos, 
NM 87545 USA}

\author{Phillip Hardee}
\affil{Department of Physics and Astronomy, the University of Alabama, 
Tuscaloosa, AL 35487 USA}

\begin{abstract}

M87 provides the best opportunity to study the base of a jet where it
is collimated and accelerated.  The size of that region scales with
the mass of the black hole, and M87 has the best combination of high
mass, proximity to the Earth, and presence of a bright jet.  VLBI
observations of M87 can probe regions under 100 gravitational radii
where theoretical studies suggest that the jet formation and
acceleration occurs.  A one-year sequence of 43 GHz observations every
3 weeks on the VLBA is being used to study the structure and dynamics
in this region.  Initial results from that effort are reported here,
including the observation of rapid motions - sufficiently rapid that
more frequent observations are planned for early 2008.  The
contribution ends with a discussion of prospects for future VLBI
observations of M87 with VSOP2.  For VSOP2, a strong recommendation is
made that a series of daily observations at 43 GHz be planned.

\end{abstract}



\section{Introduction}

Significant progress has been made in recent years in understanding the
theory of the launching of jets from accretion disk/black hole systems
\citep{Ha06,McK06,Miz06}.  Simulations are able to produce jets
starting with an accretion disk without ordered magnetic fields.  The
fields are stretched by the differential rotation of the disk, and
loops rise above, but remain attached to the disk, and ultimately the
black hole.  A structure forms along the rotation axis that can have a
relativistic Poynting jet in the center surrounded by a magnetic field
twisted into a helical shape by the disk rotation and filled by
material from a disk wind or corona.  The jet is not fully accelerated
and collimated until several hundred Schwarzschild radii (R$_s$).
Within this scale, it might be possible to see evidence for the
acceleration and collimation process and to constrain the theory.

There are very few objects in which structures in a jet of under a few
hundred R$_s$ can be observed, even with high frequency VLBI.  A
favorable combination of high mass black hole, proximity to the Earth,
and bright jet is required.  Galactic objects are close, but too low
mass except for the Galactic Center.  The black hole at the Galactic
Center has a high angular size black hole, but no observed jets.  The
best source for such a study is M87.  M87 has a super-massive black
hole with a mass of about $3\times 10^9 M_{\sun}$ \citep{harms94,
macchetto97, marconi97} and, as the dominant galaxy of the Virgo Cluster,
has a distance of only 16 Mpc \citep{whitmore95,tonry01}.  For these
parameters, the Schwarzschild radius of about 60 AU subtends 3.7
microarcseconds.  M87 has a bright jet that can be studied at the
highest frequencies, and hence, highest resolutions available to Very
Long Baseline Interferometry (VLBI).  Other galaxies that are as close
or closer either have lower mass black holes or don't have adequately
bright jets to allow detailed observations.

The basic structure on sub-parsec scales of M87 was first published by
\citet{jun99} based on 43 GHz global VLBI observations.  That
structure includes a very wide opening angle in the first few tenths
of a milli-arcsecond (mas) and a pronounced limb-brightened jet structure.
The wide opening angle suggests that the collimation region is
beginning to be resolved.  The jet has been observed with VLBI many
times since \citep[see] [and references
therein]{dodson06,kov07,ly07,kr07}.  Typically motions significantly
less than the speed of light are seen, including recent results at 15
GHz where features are almost stationary \citep{kov07}.  At lower
resolution, farther from the core, superluminal motions of $4c$ to
$6c$ have been seen, for example at the HST1 knot \citep{bi99,ch07}.
But at higher resolutions, the situation is unclear because any fast
motions would be seriously under-sampled by the previous observations.

We report here on the status of our efforts to delineate properly the
motions in M87 on scales from around 60 to several hundred R$_s$ in
projection using observations with the Very Long Baseline Array
\citep[VLBA:][]{napier94} at 43 GHz every 3 weeks for a year.  Higher
resolution could have been obtained using 86 GHz, or a global array,
but the imaging difficulties and practical scheduling constraints made
any such project far more difficult.  For the declination of M87
($12\deg$), the typical resolution of the VLBA at 43 GHz of $0.4
\times 0.2$ mas (elongated north-south) corresponds to 110 by 54 R$_s$
or 0.031 by 0.015 pc.  Note that distance along the jet is
foreshortened if the jet is near the line-of-sight as suggested by the
observation of superluminal motions.  For example, for the $\gamma =
6$ needed to see $6c$ motions, the angle to the line-of-sight is
likely to be near $10\deg$, giving a foreshortening by a factor of
about 6.

\section{Observations}

Our study of the inner jet of M87 began with a study of archival data
from 5 VLBA observations at 43 GHz spaced at intervals of roughly a
year.  The data were originally taken for various reasons, usually to
use M87 as a phase reference source for observations of a nearby weak
source.  These data were reprocessed and examined for indications of
structural evolution \citep{ly07}.  While the general character of the
structure was consistent from epoch to epoch, it only seemed possible
to identify corresponding components in more than one image by using
the two most closely spaced observations.  If those components really
were related, apparent speeds of $0.25c$ to $0.40c$ were seen.  But
any higher speeds would have made it impossible to identify components
at more than one epoch and, given the superluminal speeds seen in
other sources and elsewhere in M87, such a case could certainly not be
excluded.

The images from the archival data also revealed emission on the
opposite side of the ``core'', or brightest feature, at 3 epochs.
That feature was also seen in the 15 GHz observations of \citet{kov07}
who see it extending to about 3 mas from the core.  It is an open
question as to whether that feature is the counterjet, or the inner
jet with the radio core offset from the actual black hole location.
The facts that it is seen out to 3 mas, or about 800 R$_s$ in
projection and seems to get wider with distance from the core, make
the offset black hole possibility seem unlikely, as does the
observation of circular polarization at the core position
\citep{ho06}.  If it is the far side jet of a symmetric system, and a
speed and brightness ratio can be determined, it can give strong
evidence for the true speed and orientation to the line-of-sight of
the jet, a possibility explored in \citet{ly07}.

The archival data raised a serious question about possible
under-sampling of any motions so a project to make a fully sampled VLBA
movie at 43 GHz was begun.  The first step of that project was to
determine the proper sampling rate.  That was done using 6 observations
at spacings of between 3 and 97 days during 2006.  Based on the
results of those observations, an interval of 3 weeks was chosen for
the movie and a year-long series of 18 observations was proposed.

\begin{figure}
\plotone{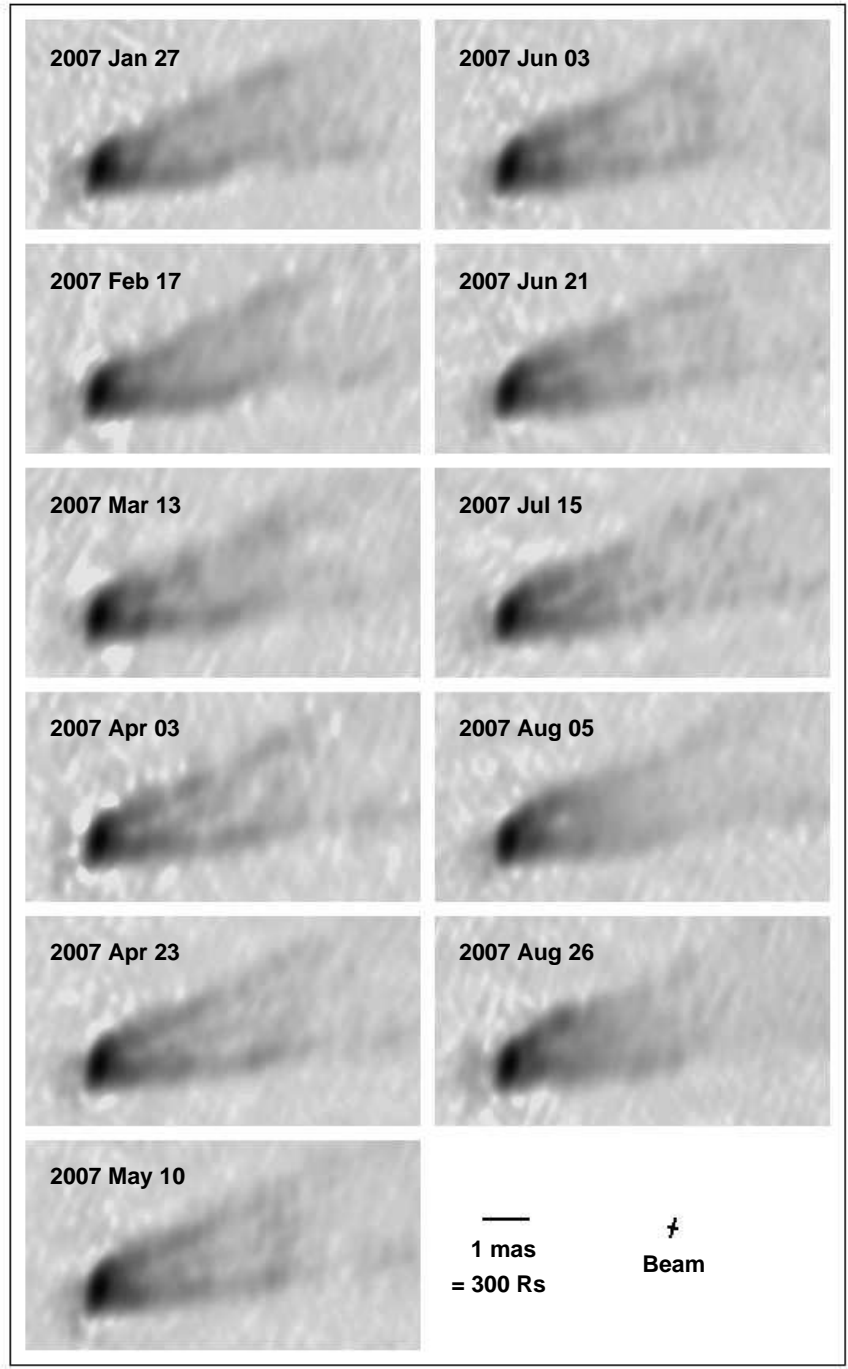}
\caption{A montage of the first 11 epoch images from the M87 image
project. The resolution (beam) is $0.43 \times 0.21$ mas elongated along
position angle $-16^\circ$ as shown by the labeled cross.  Each image
covers a region of 8.7 by 4.6 mas.}
\label{montage}
\end{figure}

\begin{figure} 
\plotone{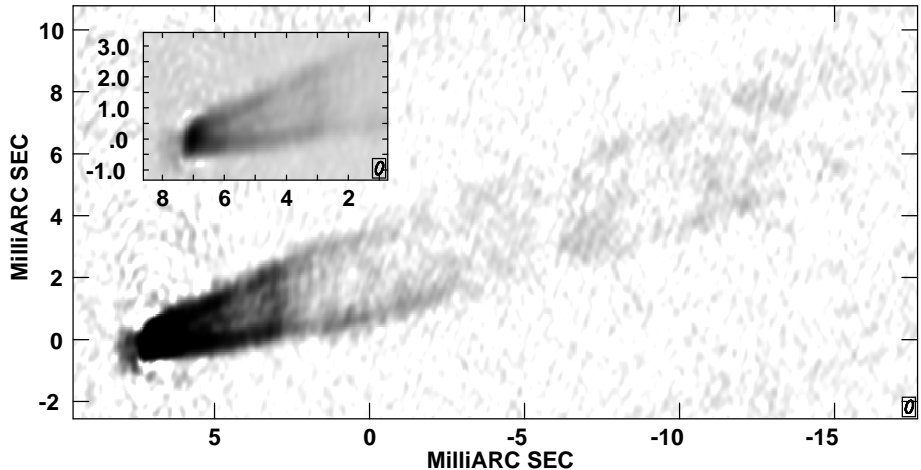} 
\caption{A composite image of M87 at 43 GHz made by summing the images
from the first 9 epochs of the movie project.  The resolution is $0.43
\times 0.21$ mas elongated along position angle $-16^\circ$.  The
insert is the central region with a transfer function that shows the
higher level emission.  The image peak is 643 mJy beam$^{-1}$ and the
off-source rms is 0.18 mJy beam$^{-1}$.  Because this image is the sum
of several images made at different times, individual features will be
blurred out and the jet will appear smoother than it actually is, much
like what is seen in a long-exposure photograph of moving water.}
\label{stack} 
\end{figure}

The movie observations began on 2007 January 27 and were still in
progress at the time of this meeting.  Images from the first 11
observations were available for the meeting.  They are shown
Figure~\ref{montage}.  The appearance of the source is generically
similar at all epochs.  The source is always a distinctly edge
brightened jet with emission filling in the central region.  The
feature on the counterjet side of the brightest emission is clearly
seen at all epochs.  In some of the epochs, there appears to be a
central spine.  The jet has structure, but that structure has more of a
lumpy but continuous nature and is not easily described as a series of
``components''.  Some features do seem to appear in multiple images,
but no very clearly defined features persist for many epochs.  To
improve the sensitivity to weak stuctures and to bring out the
persistent aspects of the structure, a weighted average image was made
from the first 9 epochs and is shown in Figure~\ref{stack}.  Note that
any detailed, changing structure is washed out in the average.

When the images are presented as a montage as in Figure~\ref{montage},
any motions are not immediately apparent.  But when assembled as a
movie and played reasonably fast, it is clear that there is overall
fast, outward motion.  It is something like watching a smoke plume
that is moving in bulk, but that also has rapid evolution of its
internal structure.  It is impossible to display such a movie here,
but one can be found at \verb+http://www.aoc.nrao.edu/~cwalker/M87+.
That website will be updated as more epochs become available.

The movie of M87 shows fast motions.  To quantify the speed in the
absence of persistent components, the difference in positions of
apparently related features in adjacent epochs were measured.  Such
differences consistently came out to be about 0.5 mas indicating a
speed of about 9 mas yr$^{-1}$ or about $2c$.  That is much faster
than motions seen in other VLBI observations of the inner regions of
this source, but those observations could not have measured such
speeds because of undersampling.  In fact, even the current
observations are undersampled because the motion is about 2.5 times
the beam width per epoch.  Once this situation was clear, the pilot
project was revisited, but any such fast motions were still not clear
in those images.  It seems a significant sequence is required for them
to be obvious.  As a result of this speed determination, a proposal
was made to extend the sequence with 10 additional observations at 5
day intervals.  Those observations are in progress as this is being
written.

The M87 43 GHz movie project is a work in progress.  But it has
already shown that the inner jet, on scales near 100 R$_s$, has mildly
relativistic motions, contrary to the conclusions of previous observations.
It has an edge brightened structure with a possible central spine and with a
counterjet side feature with motions whose character is not yet clear.  
As the observations and processing are completed, attempts will be made 
to delineate the velocity field and to constrain models of jet formation.

\section{Prospects for VSOP2}

M87 presents a prime target for VSOP2.  Ground observations with the
VLBA at 43 GHz, and with a global array at 86 GHz \citep{kr07} are
starting to resolve the wide-opening-angle base of the jet.  Another
factor of four in resolution over our VLBA results will allow
significant details to be seen in this important region and will help
guide the theoretical understanding of the structure and dynamics of
the jet launch region.  But the observations must be made frequently
enough that they are not under-sampled.  The data presented here show
motions of about 0.025 mas per day, which, with a VSOP2 beam of 0.05
mas, should be sampled daily --- much more frequently than now
envisioned.  To obtain sufficient information on the dynamics, this
observing pace must be maintained for at least a month.
This will be a major strain on observing resources, but this
is probably the best opportunity for VSOP2 to have an impact on the
understanding of the physics in the launch region of an astrophysical
jet.

\acknowledgements

This research has made use of the NASA Astrophysics Data System.
W. Junor was partially supported through NSF AST 98-03057.  C. Ly
acknowledges support through the NRAO Graduate Research Program.
P. Hardee is supported by NASA/MSFC cooperative agreement NCC8-256 and
NSF award AST-0506666 to the University of Alabama.  The National
Radio Astronomy Observatory is a facility of the National Science
Foundation, operated under cooperative agreement by Associated
Universities, Inc.


\end{document}